\documentclass[twocolumn,showpacs,preprintnumbers,
amsmath,amssymb,superscriptaddress]{revtex4}

\usepackage{graphicx}% Include figure files
\usepackage{dcolumn}% Align table columns on decimal point
\usepackage{bm}% bold math

\begin{document}

\bibliographystyle{unsrt}

\title{Neutron to proton ratios of quasiprojectile and midrapidity emission
 in \\
the $^{64}$Zn + $^{64}$Zn 
reaction at 45 MeV/nucleon}
\newcommand{\instone}{Laboratoire de Physique Nucl\'eaire,
D\'epartement de Physique, Universit\'e Laval, Qu\'ebec, Canada 
G1K 7P4.}
\newcommand{\insttwo}{Department of Chemistry and Indiana University Cyclotron Facility, 
Indiana University, Bloomington, IN 47405, USA}
\newcommand{\instthree}{Cyclotron Institute Texas A\&M University, College Station, TX 77843, USA}
\author{D. Th\'eriault \footnote{Correspondence to D. Th\'eriault, 
email: dany.theriault.1@ulaval.ca}} \affiliation{\instone}
\author{J. Gauthier} \affiliation{\instone}
\author{F. Grenier} \affiliation{\instone}
\author{F. Moisan} \affiliation{\instone}
\author{C. St-Pierre} \affiliation{\instone}
\author{R. Roy} \affiliation{\instone}
\author{B. Davin} \affiliation{\insttwo}
\author{S. Hudan} \affiliation{\insttwo}
\author{T. Paduszynski} \affiliation{\insttwo}
\author{R.T. de Souza} \affiliation{\insttwo}
\author{E. Bell} \affiliation{\instthree}
\author{J. Garey} \affiliation{\instthree}
\author{J. Iglio} \affiliation{\instthree}
\author{A.L. Keksis} \affiliation{\instthree}
\author{S. Parketon} \affiliation{\instthree}
\author{C. Richers} \affiliation{\instthree}
\author{D.V. Shetty} \affiliation{\instthree}
\author{S.N. Soisson} \affiliation{\instthree}
\author{G.A. Souliotis} \affiliation{\instthree}
\author{B.C. Stein} \affiliation{\instthree}
\author{S.J. Yennello} \affiliation{\instthree}

\date{\today}

\begin{abstract}
Simultaneous measurement of both neutrons and charged particles emitted in the
reaction $^{64}$Zn + $^{64}$Zn at 45 MeV/nucleon allows comparison of the
neutron to proton ratio at midrapidity with that at projectile rapidity.
The evolution of N/Z in both rapidity regimes with increasing centrality is
examined. For the completely re-constructed midrapidity material one finds 
that the neutron-to-proton ratio is above that of the overall 
$^{64}$Zn + $^{64}$Zn system. In contrast, the re-constructed ratio 
for the quasiprojectile is below that of the overall system. This difference
provides the most complete evidence to date of 
neutron enrichment of midrapidity nuclear matter 
at the expense of the quasiprojectile.\\ 
\end{abstract}

\maketitle

The density dependence of the asymmetry term for nuclear matter is a topic of
considerable interest \cite{mul95,li01}. 
Based on thermodynamic considerations it is possible 
for the binary nuclear fluid to fractionate into a proton-rich high density 
phase and a neutron-rich low density phase \cite{mul95}. A manifestation
of the density 
dependence of the asymmetry term 
from a kinetic perspective would involve 
the preferential transport of neutrons as compared to protons
(``isospin diffusion'') when two heavy nuclei collide \cite{shi03,li04,bar05}. 
The overlapping tails of the two colliding nuclei 
leads to a low-density region into 
which the preferential flux of neutrons may occur. 
Neutron enrichment of the low density phase 
at midrapidity in 
comparison to the high density phase at the projectile and target rapidity
might be a result of the density dependence of the 
asymmetry term. 
Following the contact phase of the collision, 
larger surface-to-volume ratio for transiently deformed reaction 
partners may also result in the preferential emission of neutron-rich clusters 
in the direction of the contact \cite{hud04,hud06}. Experimental results 
\cite{dem96,pla99,lar00,mil01,she03,hud05} show that light particles and fragments emitted 
at midrapidity exhibit a neutron enrichment compared to the quasi-projectile (QP) 
or quasi-target (QT) in mid-peripheral collisions. Results 
with free-neutron detection aiming to reconstruct 
completely the midrapidity material (MRM) \cite{sob00} have been obtained in the reaction 
$^{129}$Xe+$^{nat}$Sn at 40 MeV/nucleon. The conclusions of the authors were 
that, on average, the MRM and the bulk system have 
an indistinguishable value of their N/Z ratios \cite{sob00}. In a recent 
analysis, also with free-neutron detection, results obtained for the $^{58}$Ni+$^{58}$Ni at 
52 MeV/nucleon \cite{the05} show that the N/Z ratio  
of the QP is below the ratio of the initial system and that the MRM ratio could be above 
the ratio of the initial system. Both experiments used different experimental setups 
to match neutron and charged particle data. In the present article, we report on an 
experimental evaluation of the MRM and QP N/Z ratios made using a single setup which 
should allow a more precise measurement of the N/Z ratios.

To examine the potential neutron enrichment of midrapidity nuclear matter 
we elected to 
study a symmetric projectile-target system thus eliminating any initial
driving force towards N/Z equilibration. 
The experiment was performed at the Cyclotron Institute of Texas
A\&M University where 
a $^{64}$Zn beam, accelerated to 45 MeV/nucleon, bombarded
a self-supporting  $5$  mg/cm$^2$ $^{64}$Zn target. Charged reaction 
products were detected with the FIRST array, consisting of three annular 
telescopes denoted T1 (270$\mu$m Si(IP)-1mm Si(IP)-CsI(Tl)/PD), 
T2 (300$\mu$m Si(IP)-CsI(Tl)/PD), and T3 
(300$\mu$m Si(IP)-CsI(Tl)/PD). 
These telescopes subtended angles 
from 2.07$^\circ$ to 27.48$^\circ$ in the laboratory. 
Isotopic resolution was achieved 
for ions up to Z=12 in T1, Z=8 in T2, and Z=7 in T3 \cite{pad05}. 
Charged particles emitted at larger angles 
(36.38$^\circ$$\le$$\theta_{lab}$$<$ 51.11$^\circ$) 
were detected by  LASSA telescopes which provided isotopic resolution 
for Z$\le$6 \cite{dav01}.
Charge identification 
starts at Z=3 for T1, Z=2 for T2 and Z=1 for T3. The electronic trigger 
required the detection of at least one particle in FIRST. 
To detect emitted neutrons, 7 Bicron BC-501A liquid 
scintillator cells readout by photomultiplier tubes 
were mounted outside the scattering chamber 
at polar angles of 
27$^\circ$, 35$^\circ$, 50$^\circ$, 63$^\circ$, 75$^\circ$, 105$^\circ$, 
145$^\circ$. Neutron energies 
were determined by time-of-flight measurements and spectra were corrected for 
background emission using a shadow bar technique. Energetic protons at 
forward angles were rejected using thin scintillator paddles placed 
directly in front
of the neutron detectors. The neutron efficiency of the experimental setup  
was determined 
by using GEANT4 \cite{ago03} simulations.\\
   
We focus on peripheral and mid-peripheral collisions 
in which a fragment with Z$\ge$9 and a parallel 
velocity (beam direction)  
larger than the center of mass velocity (4.5 cm/ns) 
was detected. The most likely origin of this large atomic number,
Z$_{res}$, high-velocity 
fragment is the decay 
residue of the QP following a non-central collision. 
Due to the 
experimental acceptance the corresponding residue of the QT 
is not detected. Nevertheless, 
due to the symmetry of the system the average properties of the 
QP and the QT are inferred to be the same.
In addition, a multiplicity of 4 charged particles and a total detected charge 
of 25 or more, Z$_{tot}\ge$25 (Z$_{BEAM}$=30), 
were also required to ensure sufficient reconstruction of the QP. 
Non-isotopically resolved ions in FIRST 
were assigned an average mass 
based on the initial N/Z of the system for Z$\ge$29
or on the evaporation attractor line \cite{cha98}. 

To compare the N/Z of the QP to that of the MRM we reconstructed the QP 
based on the measured residue, bound clusters (A$\ge$2), and free nucleons.
For bound clusters, the QP source reconstruction procedure relies on three 
hypotheses \cite{gin98}: 
1) the heaviest fragment 
selected as the QP evaporation residue has a velocity close to the QP velocity, 
2) the 
QP emission is isotropic in its reference frame and 
3) all the particles with a 
parallel velocity greater than that of the residue (forward distribution) were 
emitted from the QP. 

\begin{figure}
\includegraphics[height=13cm]{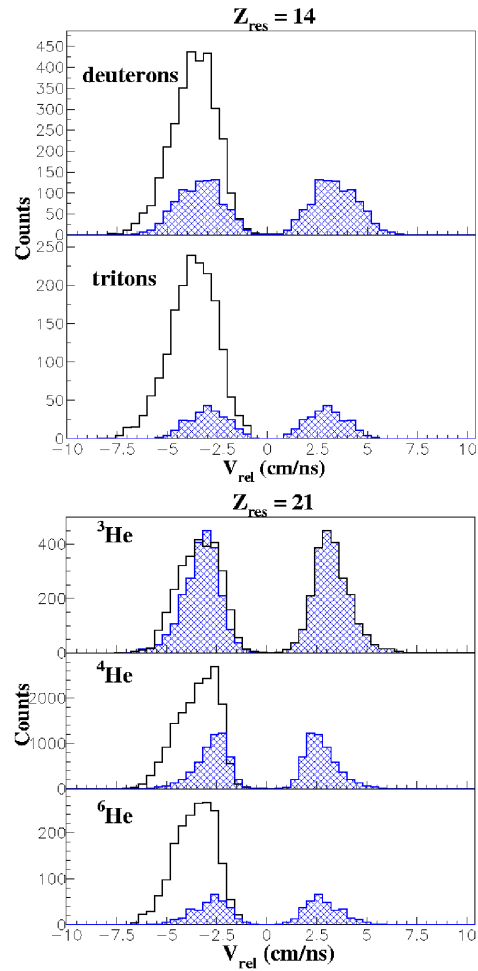}
\caption{\label{fig:fig1}(Color online) Total relative velocity between a cluster
and the QP residue, V$_{rel}$, backward distribution (clear) and 
forward V$_{rel}$ distribution backward-reflected (shadowed) for Z=1
isotopes coupled with a QP residue of Z=14 (upper panel) and for Z=2 isotopes 
coupled with a QP residue of Z=21 (lower panel).}
\end{figure}

Clusters with a minimum parallel  velocity (to minimize the 
contribution originating from
the QT) are attributed to the QP or MRM based on 
their relative velocity with the QP.
For each cluster-QP residue pair, 
the norm of the relative velocity between a cluster 
and the residue, V$_{rel}$,  is calculated.
The resulting spectra of V$_{rel}$ for hydrogen and helium isotopes 
with $Z_{res}=14$ and $Z_{res}=21$ 
are displayed in Fig. \ref{fig:fig1}.
Evident from these spectra is the fact that backward emission 
(V$_{rel}$ $<$0) is favored over forward emission (V$_{rel}$ $>$0). This 
result is consistent with previous work \cite{mon94,pla99,lef00}. To 
facilitate this comparison the distribution for V$_{rel}$ $>$0 has been 
reflected, consistent with the isotropic decay hypothesis of the QP. A striking
feature of these spectra is that the enhancement of yield for 
V$_{rel}$ $<$ 0 is particularly evident for neutron rich clusters such as
$^6$He as compared to $^3$He \cite{dem96,hud05}.
Deuterons and tritons also exhibit a 
strong preference for the backward direction.
Based upon such experimental V$_{rel}$ spectra, 
probability tables for attributing a 
cluster to the QP are constructed. 
For particles emitted forward of the QP residue the attribution probability
is taken to be unity while the probability for 
backward emitted particles 
(parallel velocity less than that of the QP residue) the attribution 
probability is
obtained by dividing the forward V$_{rel}$ distribution
by the backward distribution. 
The attribution probability, for each cluster, 
obtained in this manner depends on both 
V$_{rel}$ and $Z_{res}$ and is applied on an
event-by-event basis. 
This approach to re-construct the 
QP has been previously applied to simulations of the collision (DIT+GEMINI 
\cite{cha88,tas91}) 
and has been shown to correctly attribute more than 85$\%$ of 
the particles emitted in mid-peripheral 
and peripheral collisions \cite{gin98}.  
Further details about the reconstruction method and its efficiency are 
given in Ref. \cite{gin98,he00,gin02,he02}. Backward emitted particles 
that were not assigned to the QP belong to the MRM if their parallel velocity 
is greater than 2 cm/ns in the reference frame of the QT, which is moving 
at velocities determined in the proton multisource analysis.\\

\begin{figure}
\includegraphics[height=13cm]{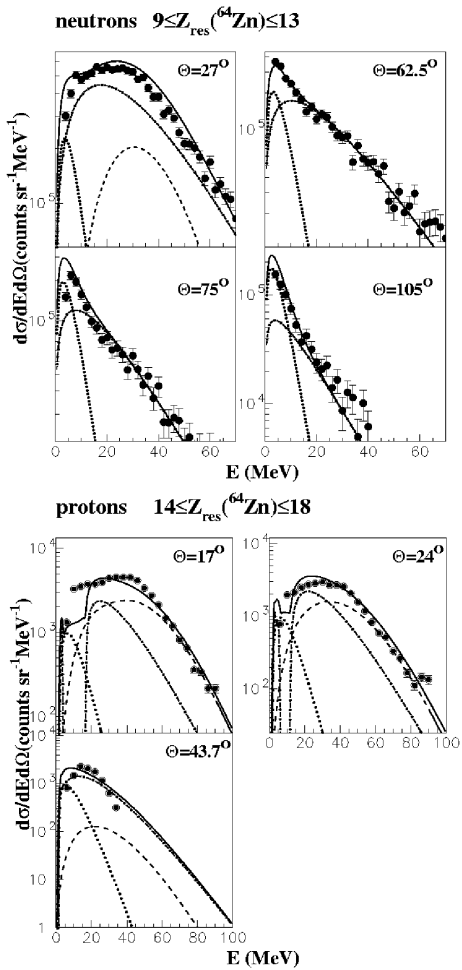}
\caption{\label{fig:fig2}Multisource fit for free neutrons (upper panel)
 and free protons (lower panel) at selected angles. Selected centrality 
classes defined with the charge of the QP residue are shown. QP 
(dashed lines), QT (dotted lines), MRM (dotted-dashed lines) and total 
(full lines) contributions are illustrated.}
\end{figure}

In order to attribute protons and neutrons to the QP and MRM,
the average proton and neutron multiplicities were 
extracted for three classes of events namely 
Z$_{res}$ = $9-13$, $14 -18$ and $19-26$. 
It has been previously shown that Z$_{res}$ is 
correlated to the excitation energy per nucleon of the QP and is thus 
a reasonable measure of the centrality of 
the collision \cite{dor00,lan01}.   
Proton multiplicities are 
obtained {\it via} a moving source analysis of energy spectra at 
$\theta_{lab}$=17$^\circ$, 24$^\circ$, and 43.7$^\circ$ while  
neutron multiplicities were extracted from a moving source 
analysis of the background and efficiency corrected energy spectra at 
7 angles. Energy spectra at all angles are fitted simultaneously 
with the emission from 
three sources, QP, QT and a midrapidity ``source'', although the
neutron spectra and proton spectra are treated independently. 
The midrapidity source, though perhaps not a single physical statistical source,
is necessary to describe the dynamical nucleon emission at midrapidity
that contributes to the N/Z of the MRM.
Each source is 
characterized by a temperature parameter $T$, 
velocity $V_s$, Coulomb barrier $B_{c}$ (protons) 
and a normalization factor $N$, related to its multiplicity. 
The nucleon energy (E) distributions are fitted under the 
assumption of surface 
emission for the QT and QP sources  and volume emission for the 
MRM by \cite{gol78,wad89,lar99,dor00,lan01}:
\begin{equation}
\label{equa:equa1}
 (\frac{d^{2}\sigma }{d\Omega dE})= 
\frac{N}{kT^i}(E-B_{c})^j\exp [-(E-B_{c})/T]
\end{equation}

where (i,j,k)=(2,1,4$\pi$) in the case of surface emission and (i,j,k)=(3/2,1/2,2$\pi^{3/2}$) in the 
case of volume emission. 
Based on the symmetry of the system, 
it is assumed that QP and QT have the same multiplicity, temperature, 
and Coulomb barrier while the velocity of the MRM is fixed at the 
nucleon-nucleon center of mass velocity (4.5 cm/ns). 
The spectra associated with the three sources is transformed to the 
laboratory frame. Fig. \ref{fig:fig2} presents the results of the 
multisource fits for protons and neutrons at selected centrality classes 
and angles. As can be seen in these representative spectra 
the multi-source fits
provide a reasonably good description of the measured energy spectra. The 
somewhat poor description of the high energy tail for the proton spectra at
$\theta_{lab}$= 43.7$^\circ$ is due to a decreased triggering efficiency
for fast protons in the LASSA telescopes. However, the uncertainty 
associated with the limited experimental data does not significantly effect 
the multiplicity associated with the QP and MRM. The 
parameters extracted from the multisource fits are presented in Table 1.

\begin{table*}
\begin{tabular}{ccccccc}

Z$_{res}$ & T$_{MRM}$ (n/p)  & T$_{QP}$ (n/p) & V$_{QP}$ (n/p)& B$_{c}$ & M$_n$ (QP/MRM) & M$_p$ (QP/MRM)\\
9-13&10.9/7.7&2.7/4.3&8.1/7.8&0.06&0.95/3.25&0.93/1.73 \\
14-18&10.7/6.8&2.0/3.7&8.4/8.1&0.38&0.84/2.95&1.01/1.96 \\
19-26&10.5/6.5&2.2/2.7&8.4/8.4&1.00&0.81/2.22&1.51/1.47 \\
\hline

\end{tabular}
\caption{Source parameters (T(MeV), V(cm/ns), B$_{c}$(MeV), M(nucleons/event))  extracted from the multi-source analysis}
\end{table*}

\begin{figure}
\includegraphics[height=11cm]{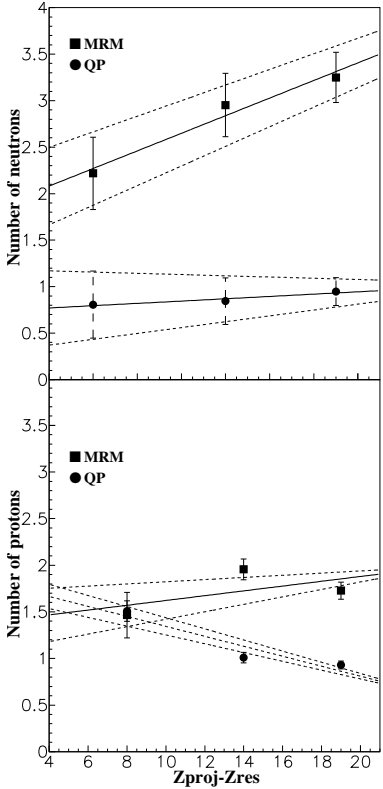}
\caption{\label{fig:fig3}Free neutron (upper panel) and 
proton (lower panel) average multiplicities as a 
function of the projectile charge minus the average QP residue charge. 
QP (dots) and MRM (squares) multiplicities are illustrated. 
Solid lines are fitted to the obtained values and dashed lines to the 
upper and lower uncertainties.}
\end{figure}

\begin{figure}
\includegraphics[height=13cm]{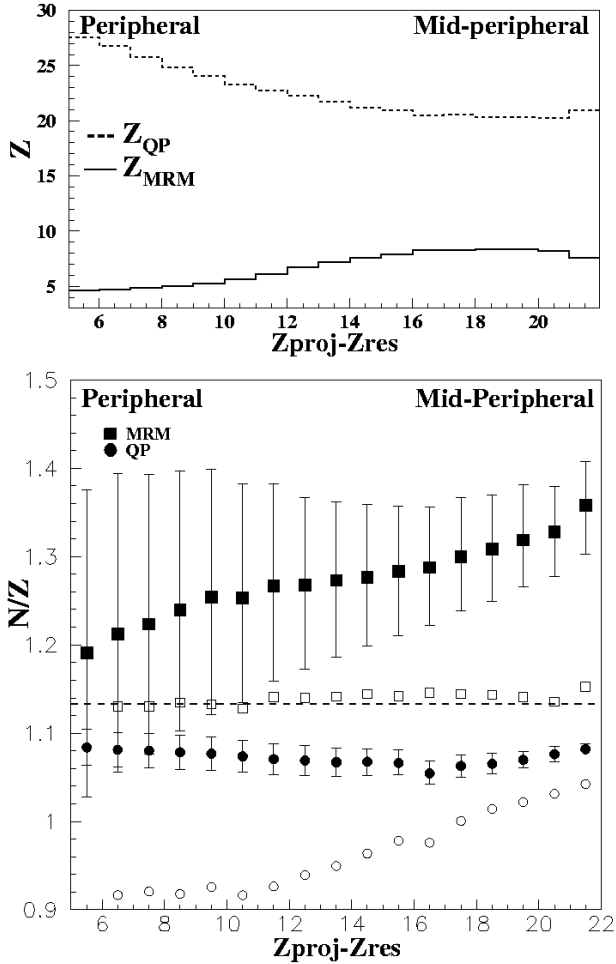}
\caption{\label{fig:fig4}Upper panel: Average charge of the QP and MRM 
as a function of the charge difference between the projectile and the residue.
Lower panel: Average N/Z ratios of QP (solid circles) and MRM (solid squares) 
as a function of the charge difference between the projectile and the residue.
The open symbols correspond to the clusters. 
Thin dotted line: original N/Z of the projectile and target. 
Errors due to multisource fit uncertainty 
for free nucleons are reported.}
\end{figure}

In Fig.~\ref{fig:fig3}, the dependence of the extracted 
proton and neutron average multiplicities on the quantity (Z$_{proj}$ - Z$_{res}$)
is shown.
This difference in atomic number is used as a measure of the centrality 
of the collision, with 
increasing (Z$_{proj}$ - Z$_{res}$) associated with
increasing centrality. 
Evident in Fig.~\ref{fig:fig3} is the slight
increase (decrease) of the proton multiplicity for the MRM (QP) with 
increasing centrality.
Measured MRM 
proton multiplicities are between 1.47 and 1.96 protons/event while QP 
proton multiplicities range from 1.51 to 0.93 protons/event. 
QP neutron multiplicities 
are between 0.81 and 0.95 neutron/event and are nearly equal to QP proton 
multiplicities for the most central collisions studied, a result 
consistent with the statistical decay of a nearly N=Z QP
at high excitation. In contrast, the MRM 
neutron multiplicity 
increases from 2.22 to 3.25 neutrons/event with increasing centrality 
and is larger than MRM proton multiplicity. 
The QP is significantly larger in size (see Fig.~\ref{fig:fig4}) 
and presumably at near saturation density.
Coulomb barrier suppresses QP statistical proton 
emission as compared to neutron emission from the QP.
Nevertheless, the neutron multiplicity extracted for the MRM 
is two to three times larger
than for the one associated with the QP. The extracted multiplicities 
were linearized as shown in  
Fig. \ref{fig:fig3} in order to provide a continuous relation between the
multiplicities of free nucleons and centrality. Error bars extracted 
from the optimization are shown.   

Evident in the top panel of Fig. \ref{fig:fig4} is the dependence of the
atomic number of the reconstructed QP and MRM 
(free nucleons, particles and all heavier fragments) on centrality. With increasing
centrality the size of the QP decreases 
from Z$\approx$28 to Z$\approx$21
while the size of the MRM increases from a value of Z$\approx$5 to Z$\approx$8.
The decrease in the size of the QP is associated with an increase in the
emitted charge. For peripheral collisions, Z$_{res}$=25 and Z$_{emitted}$=3
while for mid-peripheral collisions
Z$_{res}$=9 and Z$_{emitted}$=12, consistent with increased excitation.
For the most central collisions presented, the atomic number of the QP and
MRM are constant.

The average N/Z of the reconstructed QP and MRM 
(free nucleons, particles and all heavier fragments) are displayed in the lower panel 
of Fig.~\ref{fig:fig4}. A clusterized charge of at least 2 units 
is required in the MRM to 
avoid MRM N/Z ratios computed only with free nucleons. 
The value of the N/Z ratio for the MRM (solid squares)
increases with increasing centrality from 1.19 (peripheral) 
to 1.35 (semi-peripheral). 
For reference the N/Z of the system (1.13) is
indicated as the dashed line. Over the entire centrality range examined, the
N/Z of the MRM exceeds the N/Z of the system.
In contrast, the N/Z of the QP (solid circles)
is relatively constant over the same centrality range with a value 
of $\approx$ 1.07 $\pm$ 0.04, slightly 
less than the N/Z ratio of the original system. 
The uncertainty in the reconstructed N/Z lies largely with the uncertainty 
in the nucleon emission. To assess this uncertainty we used extreme 
assumptions regarding the contribution of free nucleon emission to the N/Z of the 
QP and MRM consistent with the measured multiplicities displayed in 
Fig. \ref{fig:fig3}. The maximum (minimum) neutron multiplicity possible for a 
particular centrality was assumed to be associated with the minimum (maximum) proton
multiplicity resulting in the largest (smallest) N/Z.
The uncertainties displayed therefore correspond to the maximum uncertainties
on the N/Z due to the free nucleon multiplicities.
Despite the large uncertainties associated with the N/Z for the MRM, 
for mid-peripheral collisions, 
a clear neutron enrichment is observed at
the expense of the QP.  
In addition to  
the uncertainty associated with the free nucleon multiplicities, 
tests were conducted 
to evaluate the uncertainty associated with the minimal 
clusterized charge required in the MRM, the mass of the 
non-isotopically resolved fragments, and also the Coulomb 
influence of the target on the 
particles with differing N/Z. 
In all these tests, a neutron enrichment of the MRM is observed 
and the calculated MRM N/Z ratio is between 1.29 and 1.37 for the 
most central collisions studied 
here.
 
To understand the observed trends in N/Z for the QP and MRM, we
have further examined the N/Z associated with 
clusters (A$\ge$2). In the case of the MRM (open squares), the N/Z is
constant with centrality and has a value close to that of the N/Z of the 
system. In contrast, for the case of the QP (open circles), 
where we additionally exclude the QP 
residue from the calculation of the N/Z, a centrality dependence 
is observed. While for the most peripheral collisions, the N/Z 
of the clusters attributed to the QP is
approximately 0.9, with increasing centrality this value increases towards
1.05. These observed trends in N/Z associated with clusters can 
be understood quite simply. 
In the case of the MRM, the tendency to clusterize at low density 
is strongly driven
by the alpha particle formation and to a lesser extent deuteron formation. 
Formation of these N=Z clusters dominates all other clusters. Formation of 
other neutron-rich clusters, such as tritons and $^6$He, explains why the 
average value of N/Z is larger than unity. Examples of the relative 
yield of these light clusters is apparent in Fig. \ref{fig:fig1}. In the
case of the QP, the measured N/Z associated with clusters is primarily driven
by the ratio of $^3$He and $^4$He at low excitation and the emission of 
heavier neutron-rich clusters with increasing excitation. 

Consequently, the increase in the N/Z for the MRM stems largely from the
preferential free neutrons as compared to free protons and clusters in this regime.
One can therefore conclude from
Fig. \ref{fig:fig4} that, on average, the QP, 
and presumably the QT, 
preferentially transfer a few neutrons to the MRM. 
While the small number of 
neutrons transferred does not change the QP N/Z significantly because 
of it's large size (21$\le$\enspace$<$Z$_{QP}$$>$\enspace$\le$28), 
it is a significant change for the much smaller MRM 
(5$\le$\enspace$<$Z$_{MRM}$$>$\enspace$\le$8). 
For the presently studied symmetric system, the Coulomb 
push of emitted particles by the QP and QT should be roughly equal 
at midrapidity thus contributing negligibly to a 
systematic suppression of neutron-rich clusters in the MRM. 
Neutron skin effects, for elements of the size 
of the $^{64}$Zn, should also be negligible \cite{sob00a}.

In summary, we observed that the average N/Z ratio of MRM is above the 
N/Z ratio of the original system for mid-peripheral 
reactions of the $^{64}$Zn+$^{64}$Zn system at 45 MeV/nucleon. This neutron 
enrichment of midrapidity nuclear material is at the expense 
of the QP that systematically has a N/Z ratio below that 
of the system. A likely origin for the preferential transfer of 
neutrons towards midrapidity 
is a density dependence of the symmetry 
energy\cite{bar98,zha05}.

\begin{acknowledgments}
This work was supported in part by the Natural Sciences and Engineering 
Research Council of Canada, the Fonds pour la Formation de Chercheurs et 
l'Aide \`a la Recherche du Qu\'ebec, 
the U.S. Department of Energy through 
Grants No. DE-FG-92ER40714 (IU), DE-FG03-93ER40773 (TAMU) and
the Robert A. Welch Foundation through 
Grant No. A-1266.
\end{acknowledgments}
\bibliography{article}
\end{document}